\documentclass[10pt,twocolumns,twoside]{IEEEtran}

\usepackage[
    top    = 54pt,
    bottom = 54pt,
    left   = 54pt,
    right  = 54pt]{geometry}
\usepackage{amssymb,amsmath,accents,amsfonts,amsthm}
\usepackage{algorithm,algorithmicx,algpseudocode,setspace}
\usepackage{tikz, verbatim, color, tabularx, booktabs, graphicx, url}
\usepackage[font=small]{caption}
\usetikzlibrary{shapes,arrows,positioning}
\usepackage[outdir=./]{epstopdf}
\usepackage{url,subcaption}
\usepackage[symbol]{footmisc}
\usepackage{mathtools}
\mathtoolsset{showonlyrefs}

\newtheoremstyle{break}
   {\topsep}{\topsep}
   {\itshape}{}
   {\bfseries}{} 
\theoremstyle{break}

\newtheorem{theorem}{Theorem}
\newtheorem{definition}[theorem]{Definition}
\newtheorem{lemma}[theorem]{Lemma}

\newtheorem{proposition}[theorem]{Proposition}
\newtheorem{corollary}[theorem]{Corollary}

\newtheorem{problem}{Problem}

\algnewcommand\algorithmicswitch{\textbf{switch}}
\algnewcommand\algorithmiccase{\textbf{case}}
\algdef{SE}[SWITCH]{Switch}{EndSwitch}[1]{\algorithmicswitch\ #1}{\algorithmicend\ \algorithmicswitch}
\algdef{SE}[CASE]{Case}{EndCase}[1]{\algorithmiccase\ #1}{\algorithmicend\ \algorithmiccase}
\algtext*{EndCase}

\newcommand{\reals}{\mathbb{R}}											
\newcommand{\prob}{\mathbb{P}}											
\newcommand{\expe}{\mathbb{E}}											
\newcommand{\todo}[1]{}													

\newcommand{\introSymbols}[1]{#1}
\renewcommand{\introSymbols}[1]{}										

\title{ \textbf{Diffusing Private Data over Networks} }
\author{Fragkiskos Koufogiannis and George J. Pappas
		\thanks{Authors are with the Department of Electrical and Systems Engineering, University of Pennsylvania, PA, USA.}
		\thanks{This work was supported in part by the TerraSwarm Research Center, one of six centers supported by the STARnet phase of the Focus Center Research Program (FCRP) a Semiconductor Research Corporation program sponsored by MARCO and DARPA.}
		\thanks{This work was supported in part by NSF CNS-1505799 and the Intel-NSF Partnership for Cyber-Physical Systems Security and Privacy.}
	}

\begin{document}

\maketitle

\begin{abstract} The emergence of social and technological networks has enabled rapid sharing of data and information. This has resulted in significant privacy concerns where private information can be either leaked or inferred from public data. The problem is significantly harder for social networks where we may reveal more information to our friends than to strangers. Nonetheless, our private information can still leak to strangers as our friends are their friends and so on. In order to address this important challenge, in this paper, we present a privacy-preserving mechanism that enables private data to be diffused over a network. In particular, whenever a user wants to access another users' data, the proposed mechanism returns a differentially private response that ensures that the amount of private data leaked depends on the distance between the two users in the network. While allowing global statistics to be inferred by users acting as analysts, our mechanism guarantees that no individual user, or a group of users, can harm the privacy guarantees of any other user. We illustrate our mechanism with two examples: one on synthetic data where the users share their GPS coordinates; and one on a Facebook ego-network where a user shares her infection status. 
\end{abstract}

\section{Introduction}
In the era of social networks, individuals' profiles include an increasing amount of private information. Besides users' intention to share this information for social interaction, their private data enables systems such as location-based services and collaborative recommender engines, that is, systems that are not part of their friendship network. Therefore, although users consent to share their private data with their friends, when this is not the case, severe privacy concerns are raised.

Traditionally, these privacy concerns are mitigated by restricting access rights (e.g. on Facebook); more precisely, only users indicated as friends are granted access to each user's personal information. However, such an approach has severe limitations as follows: first, this scheme is inflexible since users cannot be partitioned into exactly two groups, i.e. friends and strangers. Instead, privacy concerns gradually increase from family members and friends, to acquaintances, and finally, strangers. Second, a scheme based on access rights keeps private information local, which limits the ability of inferring statistics of the whole network. For instance, consider network analysts who are interested in statistics over the whole population of the social network such as population density maps and epidemic monitoring. This limits the utility of the network. Hence, an alternative mechanism that allows global statistics on the whole population and respecting individuals' privacy is needed.

Frameworks for providing privacy guarantees are differential privacy \cite{dwork2013algorithmic} and information theoretic privacy \cite{rajagopalan2011smart}.  However, most of the previous approaches in both frameworks do not consider variable privacy levels in a network, where the level of privacy depends on friendship distance. Hereafter, we consider a network where users wish to share their private data under privacy guarantees, where the strength of these guarantees is quantified by the distance on the graph. Within the context of a social network, users wish to communicate accurate information with little privacy guarantees to their close friends, whereas, they desire strong privacy guarantees whenever their private data is communicated to distant areas of the network. From the network analyst's point of view, statistics over the whole network need to be possible while ensuring the privacy guarantees.

Multiple privacy-preserving frameworks that formalize privacy guarantees have appeared in the literature, e.g. \cite{rajagopalan2011smart}, \cite{dwork06}. Commonly, privacy-preserving approaches add artificial noise to the accessed private data. This noise is designed such that the resulting response conveys little information about the private data. Specifically, an information-theoretic approach \cite{rajagopalan2011smart} constrains the mutual information between the private data and the released signal. Similarly, differential privacy \cite{dwork06}, \cite{dwork2013algorithmic} requires that the statistics of the noisy response should be \textit{almost} independent of perturbations of the private data. In this work, we adopt the framework of differential privacy because of its strong privacy guarantees, yet the underlying problem can be formulated under other privacy frameworks.

Within differential privacy, an extensive family of privacy-preserving mechanisms has emerged. The application range of these mechanisms varies from solving linear problems \cite{hsu14}, \cite{han2014differentially}, distributed convex optimization \cite{hale2015differentially}, Kalman filtering \cite{le2014differentially}, and consensus that protects the network topology \cite{katewa2015protecting} to smart metering \cite{acs2011have}, \cite{koufogiannis14} and traffic flow estimation \cite{le2014real}. In particular, the problems introduced in the aforementioned line of research share a common underlying abstract problem that can be stated as follows: given the private data \introSymbols{$u\in\mathcal{U}$ }and a predefined privacy requirement\introSymbols{ $\epsilon$}, we need to design a\introSymbols{n} \introSymbols{$\epsilon$-}differentially private algorithm\introSymbols{ $Q$}, called mechanism, which accurately approximates a desired quantity\introSymbols{ $q:u \to q(u)$}. Then, a single sample\introSymbols{ $y$} from the mechanism\introSymbols{ $Qu$} is published and is used as a proxy for the exact response\introSymbols{ $q(u)$}, so,  a curious user cannot \textit{confidently} infer the original private data\introSymbols{ $u$}. Instead of considering a single privacy level\introSymbols{ $\epsilon$} and assuming that the response are publicly released, i.e. everyone receives the same response, in this paper, we consider the novel problem of assigning \textit{different privacy levels for different users}. Moreover, contrary to publishing the responses, we assume that they are securely communicated to each user. Therefore, the aforementioned works do not address the problem introduced here. Furthermore, in \cite{alaggan2015heterogeneous}, \cite{ebadi2015differential}, multi-component private data and different privacy levels for different components are considered, i.e. in a user's profile, typically, stronger privacy is required for the component representing salary compared to that of age. Contrary to previous works that focus on variable privacy for different components of one's data, our paper focuses on different privacy levels that depend on friendship status. The work closest to ours is \cite{koufogiannis2015gradual}, where the problem of relaxing the privacy level after e.g. supplementary payments to the owners of the sensitive data. Although some of the tools in \cite{koufogiannis2015gradual} are leveraged to provide a solution, here, we consider a different problem which is the problem of releasing sensitive data to multiple parties with different privacy levels and has not been studied before.

The paper is organized as follows. Section \ref{sec:problemFormulation} informally describes the problem of diffusing private data across a network, then, provides a model of the system, reviews differential privacy, and derives a formal statement of the problem. Section \ref{sec:results} introduces a composite mechanism based on a Markov stochastic process and presents low-complexity algorithmic implementations of this mechanism. We demonstrate our approach with two illustrative examples in Section \ref{sec:examples}, one on synthetic network where a user releases her GPS coordinates and one on a Facebook ego-network where a user shares her infection status.

\section{Problem Formulation} \label{sec:problemFormulation}
Here, the problem of releasing private information over networks (i.e. social networks) is formulated. First, we provide an informal description of the problem whose formal statements are presented in Problem \ref{problem:plainVersion} and Problem \ref{problem:finalVersion} in the end of this section. Let a network be represented as a graph $G=(\mathcal{V},\mathcal{E})$, where each node $i\in \mathcal{V}$ is a user and each edge $(i,j)\in \mathcal{E}$ represents a friendship relation between users $i$ and $j$. Also, we assume that each user $i$ owns a private data $u_{i}\in \mathcal{U}$, where $\mathcal{U}$ is the set of possible private data, and wishes to share their private data with the rest of the users under privacy guarantees. Specifically, user $i$ generates an approximation $y_{ij}$ of $u_{i}$ and securely communicates $y_{ij}$ to user $j$. More specifically, each user $i$ requires her data $u_{i}$ to be $\epsilon(d_{ij})$-differential privacy against user $j$ (differential privacy is overviewed in Subsection \ref{sec:privacy}), where $d_{ij}$ is a distance function $d_{ij}: \mathcal{V}\times \mathcal{V}\to \reals_{+}$ and $\epsilon:\reals_{+}\to\reals_{+}$ is a decreasing function that converts distance $d$ to a privacy level $\epsilon(d)$. Therefore, we need to design a mechanism that generates \textit{accurate}\footnote{Here, accuracy is meant in the expected mean-squared error sense.} responses $\{y_{ij}\}_{j\in \mathcal{V}}$ while satisfying \textit{different privacy constraints for different recipients} based on the distance on the network.

In order to formalize these statements as in Problem~\ref{problem:plainVersion} and, eventually, in Problem~\ref{problem:finalVersion}, we need to revisit some concepts and known results. Subsequently, modeling assumptions are presented in Subsection \ref{sec:model}, whereas differential privacy is briefly reviewed in \ref{sec:privacy}. We present a conventional approach, i.e. a scheme based on access rights in Subsection \ref{sec:baseline}, whereas Subsection \ref{sec:statement} formally presents the problem of diffusing private data over networks.

\subsection{System Model} \label{sec:model}
Consider a network represented as a graph $G$ with $|\mathcal{V}|=N$ nodes. For simplicity, we assume that the graph is undirected and unweighted, although this assumption can be removed. Each node $i\in \mathcal{V}$ represents a user and $(i,j) \in \mathcal{E}\subseteq \mathcal{V}\times \mathcal{V}$ represents the friendship relation between users $i$ and $j$. Each user owns a private data $u_{i} \in \mathcal{U}$. Typical examples of private data include:
\begin{enumerate}
	\item \textit{Timestamps:} let $u_{i}\in\reals$ be a real-valued representation of a timestamp such as date of birth, e.g. \textit{Unix time} \cite{wikiUnixTime} is a popular way of mapping timestamps to integers;
	\item \textit{Location:} let $u_{i}\in\reals^{2}$ be the GPS coordinates of the residence of an individual $i$;
	\item \textit{Binary states:} let $u_{i}\in\{0,1\}$ indicate user's $i$ status such as infected or healthy, married or single etc.
\end{enumerate}
Further, we want the severity of the privacy concerns to scale with the distance between two nodes. Typical choices for the distance function $d:\mathcal{V}\times \mathcal{V}\to\reals_{+}$ are as follows:
\begin{enumerate}
	\item \textit{Shortest path distance:} let $d_{ij}$ be the length of the shortest path connecting nodes $i$ and $j$;
	\item \textit{Resistance distance:} let $d_{ij}$ be the resistance between nodes $i$ and $j$, where the edges of graph $G$ are associated with unit resistors \cite{babic2002resistance}.
\end{enumerate}
A more extended model can incorporate additional information such as family relationships. Further, directed edges (e.g. blocked users) can be also be allowed in social network scenarios.

\subsection{Differential Privacy} \label{sec:privacy}
Differential privacy is a formal framework that provides rigorous privacy guarantees. Differentially private algorithms add noise in order to make it hard for a curious user to infer whether someone's data has been used in the computation. The dependency of this noisy response on the private data is required to be bounded, as formally stated in Definition \ref{def:differentialPrivacy}. The strength of this bound is quantified by the non-negative parameter $\epsilon\in[0,\infty)$, called \textit{privacy level}, where smaller values of $\epsilon$ imply stronger privacy guarantees. Moreover, an adjacency relation $\mathcal{A}$ is a symmetric binary relation over the set of private data $\mathcal{U}$ which includes the pairs of private data $(u,u')$ that should be rendered \textit{almost} indistinguishable. Further, a mechanism\footnote{For a set $\mathcal{T}$ and a rich-enough $\sigma$-algebra $\mathcal{T}$ on it, we denote the set of all probability measures on $(T,\mathcal{T})$ with  $\Delta(T)$. Specifically, for Euclidean spaces $T=\reals^{n}$, we consider the Borel's $\sigma$-algebra.} $Q: \mathcal{U} \to \Delta\left( \mathcal{Y} \right)$ is a randomized map from the space of private data to the space of responses.
\begin{definition}[Differential Privacy \cite{dwork2013algorithmic}] \label{def:differentialPrivacy}
Let $\epsilon>0$, $\mathcal{U}$ be the space of private data, and $\mathcal{A}\subseteq \mathcal{U}\times\mathcal{U}$ be an adjacency relation. The mechanism $Q: \mathcal{U} \to \Delta\left( \mathcal{Y} \right)$ is $\epsilon$-differential privacy if:
\begin{align}
	\prob( Qu \in \mathcal{S} ) \leq e^{\epsilon} \, \prob( Qu' \in \mathcal{S} ), \quad \text{ for all } \mathcal{S} \subseteq \mathcal{Y},
\end{align}
for all adjacent inputs $(u,u')\in\mathcal{A}$.
\end{definition}

In this work, we consider real-valued private data \mbox{$\mathcal{U}=\reals^{n}$} and the following adjacency relation:
\begin{align} \label{eqn:adjacencyRelation}
	(u,u')\in\mathcal{A}_{2} \Leftrightarrow \|u-u'\|_{2}\leq \alpha,
\end{align}
where $\alpha\in\reals_{+}$ is a small constant. Practically, adjacency relation $\mathcal{A}_{2}$ requires that, given the output of mechanism $Q$, a curious user should not be able to infer the private input $u$ within a radius of $\alpha$. A popular differentially private mechanism is the Laplace mechanism which is near-optimal \cite{wang14}, \cite{koufogiannis14}, is used as a building block for many mechanisms, and is described next.

\begin{theorem}[Laplace Mechanism \cite{dwork2013algorithmic}] \label{def:laplaceMechanism}
Consider the mechanism $Q: \reals^{n} \to \Delta\left( \reals^{n} \right)$ that adds Laplace distributed noise:
\begin{align}
Qu = u + V, \text{ where } V \sim \text{Lap}\left( \frac{\alpha}{\epsilon} \right),
\end{align}
where $\text{Lap}(b)$ has density $\prob( V=v ) = e^{-\frac{\|v\|_{2}}{b}}$. Then, mechanism $Q$ is $\epsilon$-differential private under adjacency relation~$\mathcal{A}_{2}$.
\end{theorem}

\subsection{Access Rights Scheme} \label{sec:baseline}
\begin{figure} \centering
\begin{subfigure}{.48\linewidth} \centering
	\includegraphics[width=\textwidth]{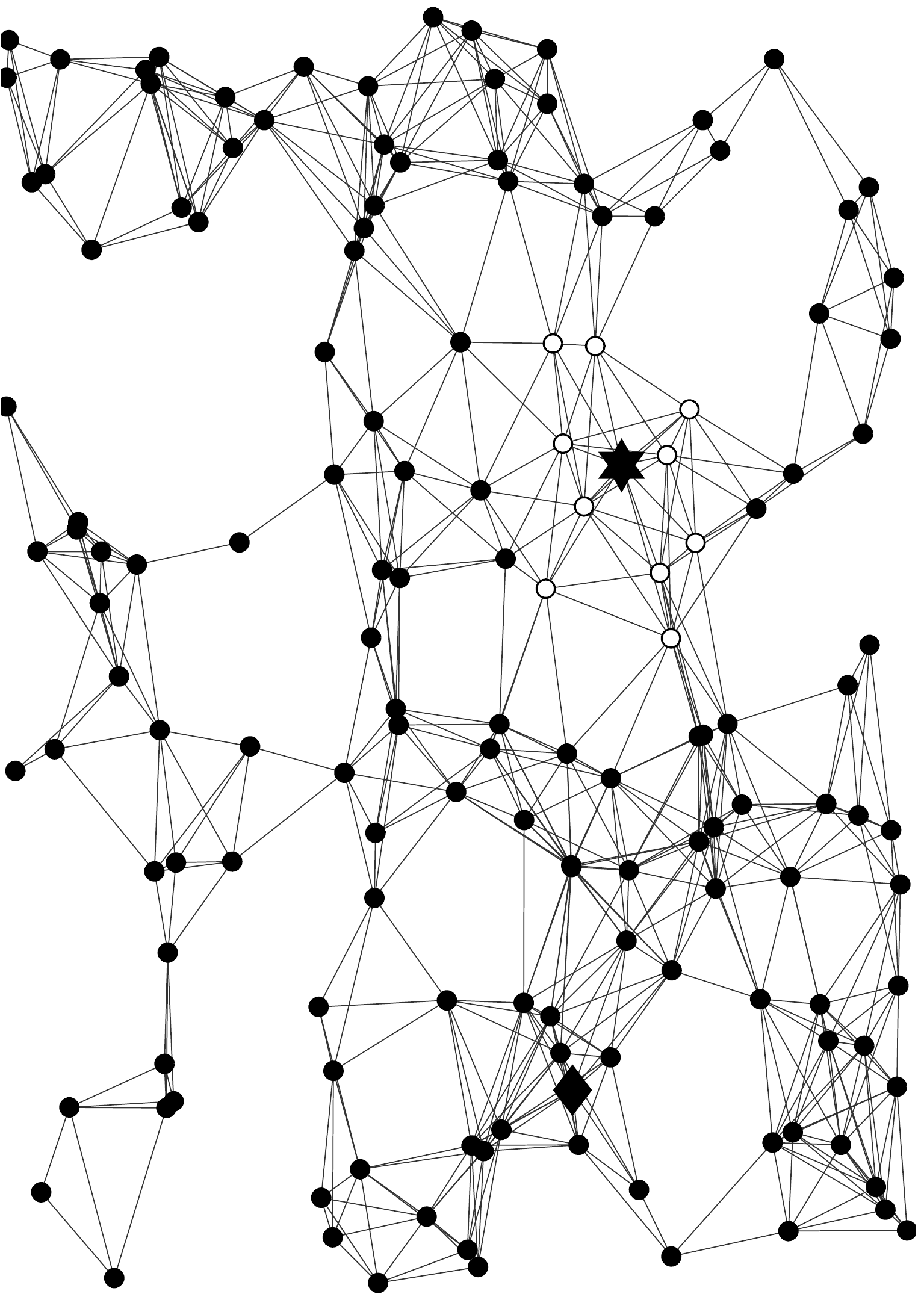}
	\caption{An access right scheme.}
	\label{fig:accessRightsNetwork}
\end{subfigure}
\begin{subfigure}{.48\linewidth}  \centering
	\includegraphics[width=\textwidth]{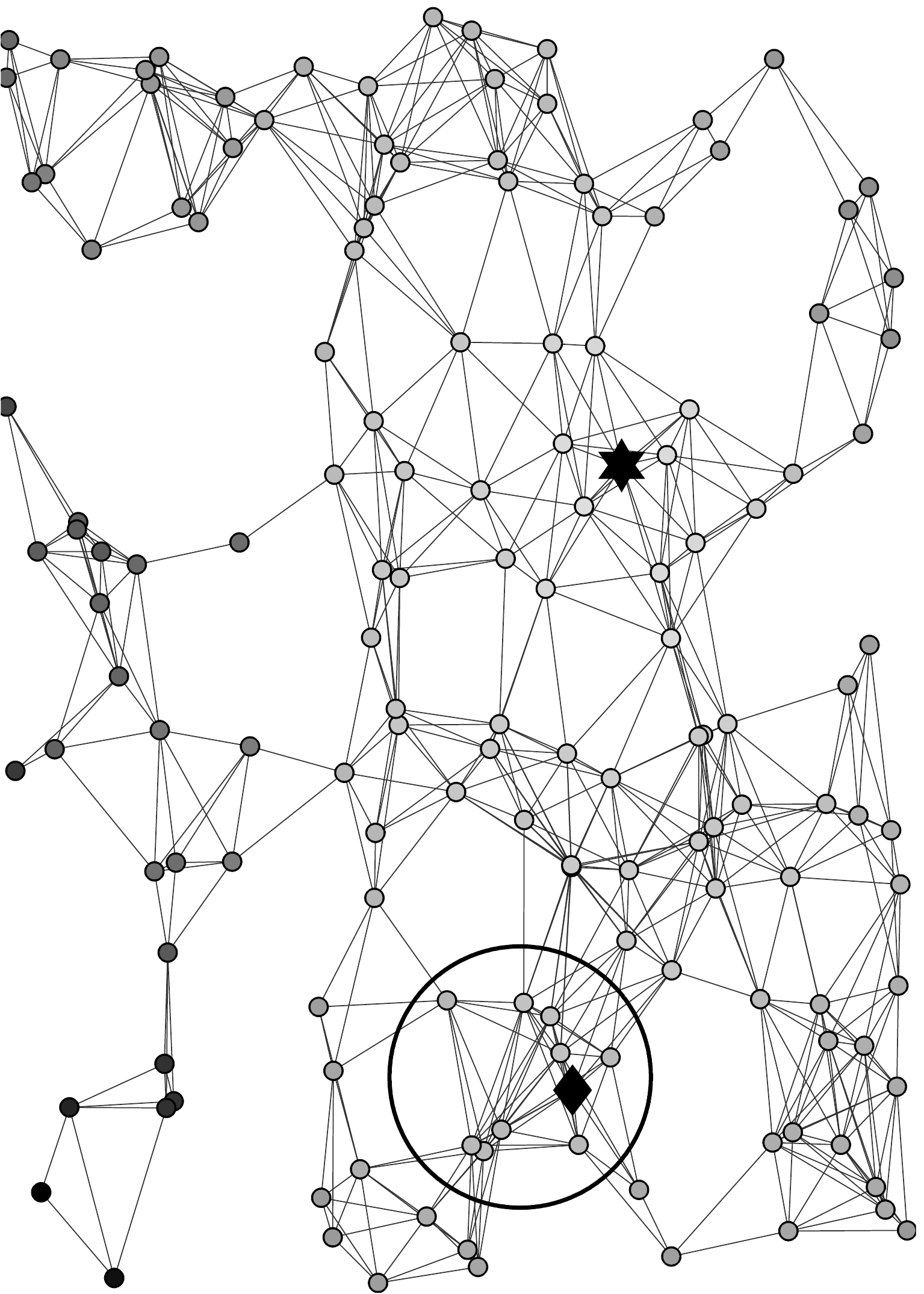}
	\caption{A distance-based scheme.}
	\label{fig:resistanceDistanceNetwork}
\end{subfigure}
\caption{A synthetic network with $150$ nodes and $1256$ edges is shown. Each node represents a user of the network and each edge indicates a friendship. The user indicated with the star wishes to share her sensitive information with the rest of the network. Privacy concerns can be addressed by managing access privileges. Under an access right scheme (Figure \ref{fig:accessRightsNetwork}), only friends of the starred user (blue nodes) are granted access to the exact information, whereas any other member (red nodes) have no access. Such a scheme partitions users to only two groups; friends and strangers. Moreover, each user has access only to local information and cannot estimate the global state of the network. Therefore, any estimator constructed by the diamond user will be independed of the data of the starred user and, thus, biased. On the other hand, Figure \ref{fig:resistanceDistanceNetwork} proposes an approach where users' privacy concerns scale with the distance from others. Friends (lighter-colored nodes) receive a less noisy versions of the private data, whereas strangers (darker-colored nodes) receive only heavily perturbed versions. Despite the increased noise, estimates of aggregate statistics are possible. However, coalitions might be encouraged and initial privacy guarantees can quickly degrade. For example, users within the circle can combine their estimates and infer the private data of the starred user.}
\end{figure}

Now, we describe a typical approach for handling privacy concerns in social network while highlighting its limitations and motivating the need for a more sophisticated privacy-aware approach. Figure~\ref{fig:accessRightsNetwork} shows a synthetic network with $150$ nodes, where the starred node wishes to share her sensitive information with the rest of the network. Privacy concerns can be handled by regulating access privileges. For example, friends of a user can access her data, whereas every other user cannot. Such a scheme has limitations. On one hand, users are coarsely partitioned to friends and strangers as depicted in Figure~\ref{fig:accessRightsNetwork}; friends of the star-labeled user are colored white whereas strangers are colored black. Instead, the distance between two users can be more finely quantified by a real-valued function. On the other hand, each user has access only to neighboring information. Although restricting access rights settles privacy concerns, computing global statistics on the network is impossible, limiting the utility of the network. Indeed, any estimator of global quantities (mean value, histogram etc.) will be biased. Therefore, users may choose to collaborate, merge their local information, and damage any privacy guarantees. Figure \ref{fig:resistanceDistanceNetwork} overcomes these limitations by defining a distance function $d:\mathcal{V}\times \mathcal{V}\to\reals_{+}$ which quantifies the strength of the privacy concerns. In this case, users share privacy-aware versions of their profile with every member of the network.

\subsection{Diffusing Sensitive Information over a Social Network} \label{sec:statement}
Under the modeling introduced in Subsection \ref{sec:model}, we pose the problem of designing a mechanism that diffuses private data over a network in the following.

\begin{problem} \label{problem:plainVersion}
Design a privacy-aware mechanism \mbox{$Q: \mathcal{U} \to \Delta\left( \mathcal{U}^{N} \right)$} that privately releases user's $i$ sensitive data $u_{i}\in\mathcal{U}$ over a social network. Specifically, design mechanism $Q$ that generates $N$ responses $\{y_{j}\}_{j=1}^{N}$, where $y_{ij}$ is the securely communicated response to user $j$. Further, for the adjacency relation \eqref{eqn:adjacencyRelation} (where, for simplicity, $\alpha=1$), the mechanism $Q$ needs to satisfy the following properties:
\begin{itemize}
	\item \textit{Variable Privacy:} The mechanism must generate the response $y_{ij}$ for private data $u_{i}$ which $\epsilon(d_{ij})$-differential private.
	\item \textit{Optimal Utility:} Response $y_{ij}$ must be an accurate approximation of the sensitive data $u_{i}$, i.e. for real-valued private data, it should minimize the expected squared-error
	\begin{align}
		\expe_{Q} \| y_{ij} - u_{i} \|_{2}^{2}.
	\end{align}
\end{itemize}
\end{problem}

Specifically, whenever individual $i$ shares her sensitive information to another individual $j$, she requires $\epsilon(d_{ij})$-differential privacy, where $\epsilon(\cdot):\reals_{+}\to\reals_{+}$ is a decreasing function that coverts a distance $d$ to a privacy level $\epsilon(d)$. People residing close (w.r.t. a distance) to individual $i$ receive a loose privacy constraint $\epsilon_{ij}\gg1$, whereas strangers get noisier versions $\epsilon_{ij}\ll1$.

Problem \ref{problem:plainVersion} admits a straightforward but unsatisfying approach. Let $y_{ij} = u_{i} + V$, where $V\sim\text{Lap}\left( \epsilon(d_{ij})^{-1} \right)$, independently for each user $j\in \mathcal{V}$. Subsequently, a group of users $j\in A\subseteq\mathcal{U}$ have the incentive to collaborate share their estimates $\{ y_{ij} \}_{j\in A}$ in order to derive a more accurate estimator $y_{A}$ of $u_{i}$ described by
\begin{align}
	y_{A} = \sum_{j\in A} w_{j} y_{ij}.
\end{align}
Figure~\ref{fig:resistanceDistanceNetwork} depicts a group of users forming such a coalition. The possibly large group $A$ resides far away from the user indicated by the star, $d_{ij} \gg 1, \forall j\in A$. Although each user $j$ in the group $A$ receives a highly noisy estimate of $u_{i}$, estimator $y_{A}$ is more accurate. The composition theorem of differential privacy \cite{dwork2013algorithmic} guarantees only $\left( \sum_{j\in A} \epsilon(d_{ij}) \right)$-privacy which can be rather looser than each of the $\epsilon(d_{ij})$-privacy guarantees; larger values of $\epsilon$ imply less privacy.

Therefore, Problem \ref{problem:plainVersion} is subject to coalition attacks. Thus, we restate Problem \ref{problem:plainVersion} by requiring that any group $A$ that exchanges their estimates $\{ y_{ij} \}_{j\in A}$ cannot produce a better estimator of $u_{i}$ than the best estimator among the group $y_{ij^{*}}$, where $j^{*} = \arg\min_{j\in A} d_{ij}$ is the user closest to user $i$. This problem can be stated as follows:

\begin{problem} \label{problem:finalVersion}
Design a privacy-aware mechanism \mbox{$Q: \mathcal{U} \to \Delta\left( \mathcal{U}^{N} \right)$} that releases a approximation of user's $i$ sensitive data $u_{i}\in\mathcal{U}$ over a social network. Specifically, mechanism $\mathcal{M}$ generates $N$ responses $\{y_{ij}\}_{j=1}^{N}$ and securely communicates response $y_{ij}$ to user $j$. Mechanism $Q$ needs to satisfy:
\begin{itemize}
	\item \textit{Privacy:} For any group of users $A\subseteq \mathcal{V}$, response $\{ y_{ij} \}_{j\in A}$ must be $\max_{j\in A} \epsilon( d_{ij} )$-differential private.
	\item \textit{Performance:} Response $y_{ij}$ must be an accurate approximation of the sensitive data $u_{i}$.
\end{itemize}
\end{problem}

\section{Main Results} \label{sec:results}
In this section, we approach the problem of diffusing private data over a network. Subsection \ref{sec:theory} derives the needed theoretical results and establishes that the accuracy of each estimate $y_{ij}$ depends \textit{only} on the distance $d_{ij}$. Moreover, algorithmic implementations of the composite mechanism $Q$ should scale for vast social networks. Subsection \ref{sec:algorithms} provides algorithmic implementations of the mechanism $Q$ with complexity $O\left( \ln \left( \frac{  \max_{i,j \in V} \epsilon(d_{ij}) }{ \min_{i,j \in V} \epsilon(d_{ij}) } \right) \right)$.

\subsection{A Private Stochastic Process} \label{sec:theory}
For $n$-dimensional real-valued private data $u\in\reals^{n}$, we derive a composite mechanism that generates the response $y_{ij}$ that user $j$ receives as an approximation of user's $i$ private data $u_{i}$. This mechanism has the following two properties. First, the accuracy of the response $y_{ij}$ depends solely on the distance $d_{ij}$ between nodes $i$ and $j$. Specifically, the expected squared-error in Equation \eqref{eqn:mse} does not depend on any other parameters of the network (e.g. size, topology) or the rest of the responses $\{y_{ik}\}_{k\in \mathcal{V} \backslash \{j\}}$. Second, any group of users $A\subseteq\mathcal{V}$ that decides to collaborate and share their responses $\{ y_{ij} \}_{j\in A}$ cannot infer anything more about user's $i$ private data. Algorithmic aspects of the composite mechanism are deferred until Subsection \ref{sec:algorithms}.
\begin{align} \label{eqn:mse}
	\expe \left( \| y_{ij} - u_{i} \|_{2}^{2} \right) = \frac{n(n+1)}{\epsilon(d_{ij})^{2}},
\end{align}
where \mbox{$\epsilon: \reals_{+} \to \reals_{+}$} is a decreasing function which converts distance $d$ to a privacy level $\epsilon(d)$. Then, Theorem~\ref{thm:socialNetworkPrivacy} introduces the underlying composite mechanism.
\begin{theorem} \label{thm:socialNetworkPrivacy}
Let $d_{ij}\in\reals_{+}$ denote the distance between users $i$ and $j$, and $u_{i}\in\reals$ be the private data of user $i$. Consider the mechanism $Q$ that generates the responses:
\begin{align}
	y_{ij} = u_{i} + V^{(i)}_{\epsilon(d_{ij})},
\end{align}
where $\{ V^{(i)}_{\epsilon} \}_{\epsilon>0}$ is a sample of a Markov stochastic process $\{ V_{\epsilon} \}_{\epsilon>0}$. Then, mechanism $Q$ provides a solution to Problem~\ref{problem:finalVersion}. In particular, it has the following properties:
\begin{itemize}
	\item The variance of response $y_{ij}$ is $n\,(n+1)\,\epsilon(d_{ij})^{-2}$ and, thus, depends only on the distance between users $i$ and $j$.
	\item For any subset of users $A\subseteq \mathcal{V}$, the mechanism that releases the responses $\{ y_{ij} \}_{j\in A}$ is $\left( \underset{ j\in \mathcal{V} }{\max} \, \epsilon(d_{ij}) \right)$-differential private.
\end{itemize}
\end{theorem}

The proof of Theorem~\ref{thm:socialNetworkPrivacy} is presented in Appendix~\ref{app:socialNetworkPrivacy}. The main idea is introducing correlation between the responses $\{ y_{ij} \}_{j\in\mathcal{V}}$. For $n=1$, the stochastic process $\{ V_{\epsilon} \}_{\epsilon}$ has closed-form expressions, whereas, for $n>1$, closed-form expressions are derived only for the infinitesimal increments \mbox{$V_{\epsilon + d\epsilon} - V_{\epsilon}$}. Nonetheless, we derive handles that allow for exact (in the sense that we do not use approximations of the process) and efficient (in the algorithmic complexity sense) sampling of the process. Furthermore, our proof techniques are robust and can possibly be applied beyond the Laplace mechanism; for example, the $K$-norm mechanism \cite{hardt10} that appears in a different setting than the one considered here.

\begin{figure}
	\begin{center}
		\includegraphics[width=\linewidth]{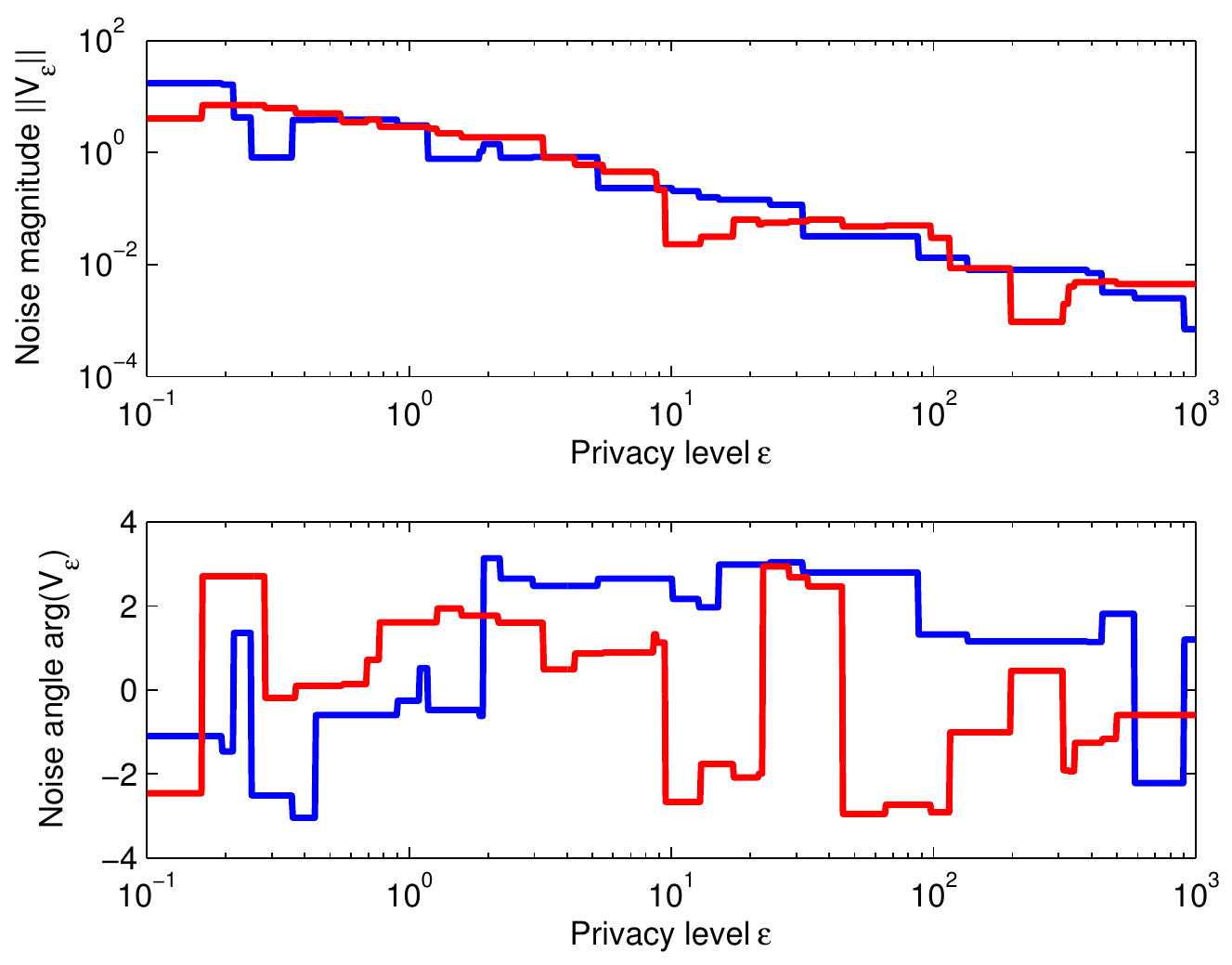}
		\caption{Two samples of the two-dimensional process which is the underlying object for diffusing private GPS coordinates over a network.} \label{fig:process2D}
	\end{center}
\end{figure}

\begin{figure}
	\begin{center}
		\includegraphics[width=\linewidth]{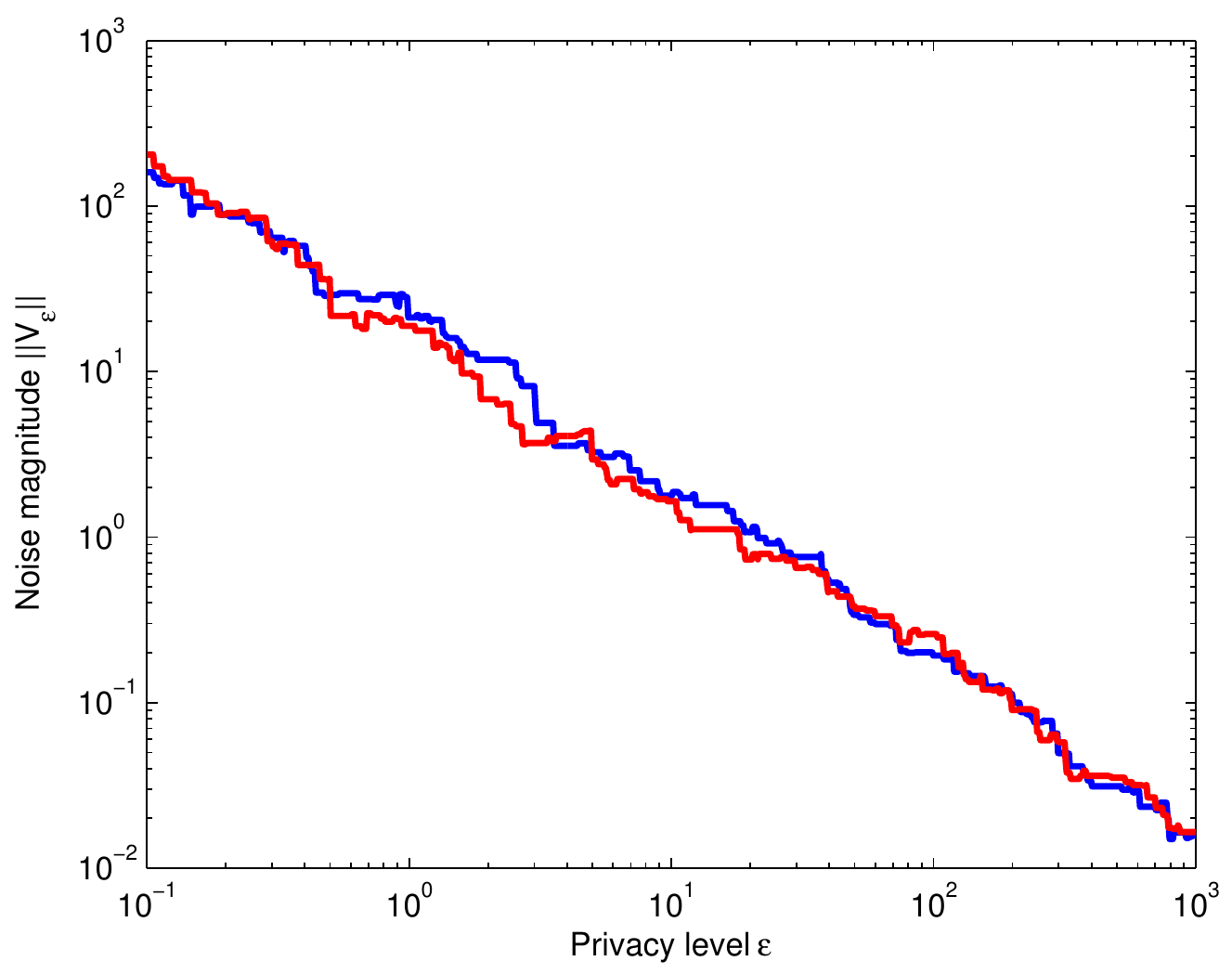}
		\caption{The $\ell_{2}$-norm of two samples of the stochastic process $\{ V_{\epsilon} \}_{\epsilon>0}$ in high-dimensions ($n=20$) which can be used to diffused private signals over networks, such as power consumption in smart grids.} \label{fig:processInfD}
	\end{center}
\end{figure}

Figure~\ref{fig:process2D} pictures two samples of the stochastic process $\{ V_{\epsilon} \}_{\epsilon>0}$, for $n=2$, in polar coordinates and shows that the process is a jump process; i.e., with high probability, the process is constant in small intervals. Figure~\ref{fig:processInfD} pictures two samples of the process in high dimensions. The process is again \text{lazy}, yet, the jumps are more often.

A major consequence of Theorem \ref{thm:socialNetworkPrivacy} is that mechanism $Q$ does not incentivize coalitions. Specifically, consider a group of curious users $A\subseteq \mathcal{V}$ who wish to estimate $u_{i}$ more accurately and, thus, collaborate and share their knowledge $\{y_{ij}\}_{j\in A}$. In practice, such a group can be fake accounts of a single but distant (in the sense of $d$) user. Then, given this shared knowledge, the best estimator is:
\begin{align}
\hat{u}_{i} = \left. y_{ij^{*}} \right|_{j^{*}\in \arg\min_{j\in A} d_{ij}}.
\end{align}
Therefore, user $j^{*}$ is not benefited by such a coalition and, thus, she has no incentive to participate in the coalition and share her information $y_{ij^{*}}$.

\subsection{Algorithmic Implementation} \label{sec:algorithms}
Sampling from a continuous-domain stochastic process can often be performed only approximately. For example, consider the Brownian motion $\{ B_{t} \}_{t\in[0,1]}$ which, for sampling purposes, requires storing an \textit{infimum} of real values. Contrary to Brownian motion, the private process $\{ V_{\epsilon} \}_{\epsilon>0}$ rarely changes value and is, thus, \textit{lazy}. More formally, restricted to a sufficiently small interval $[\epsilon_{1}, \epsilon_{2}]$, the stochastic process $\{ V_{\epsilon} \}_{\epsilon \in [\epsilon_{1},\epsilon_{2}]}$ is constant with high probability. Furthermore, assuming the existence of an algorithm for computing the distance $d_{ij}$, the response $y_{ij}$ can be generated during run-time. This property is crucial, since it circumvents the $O\left(N^{2}\right)$ memory requirements of a static implementation. Proposition~\ref{thm:privateProcessJumpStatistics} characterizes the distribution of the number of jumps in a bounded interval.

\begin{proposition} \label{thm:privateProcessJumpStatistics}
The number of jumps that the process $\{ V_{\epsilon} \}_{\epsilon>0}$ performs in the interval $[\epsilon_{1}, \epsilon_{2}]$ is Poisson distributed with mean value $(n+1) \ln \left( \frac{ \epsilon_{2} }{ \epsilon_{1} } \right)$.
	\begin{align} \label{eqn:privateProcessJumpStatistics:1}
		\prob \left( k \text{ jumps in } [\epsilon_{1}, \epsilon_{2}] \right) = \frac{ x^{k} }{k!} e^{-x},
	\end{align}
	where $x = (n+1) \ln \left( \frac{\epsilon_{2}}{\epsilon_{1}} \right)$.
\end{proposition}

\begin{corollary}
Process $\{ V_{\epsilon}\}_{\epsilon>0}$ performs $\expe[k] = (n+1)\ln2$ jumps (in expectation, with variance $\text{Var}[k]=(n+1)\ln2$) for every doubling of the privacy level, i.e. in the interval $[\epsilon,2\epsilon]$.
\end{corollary}

This laziness renders samples from the process highly-compressible. Indeed, given the locations $\{ \epsilon^{(i)} \}_{i=1}^{k}$ of the jumps and the values\footnote{We use the notation $V_{\epsilon_{-}} = \lim_{\tau\uparrow \epsilon} V_{\tau}$ and $V_{\epsilon_{+}} = \lim_{\tau\downarrow \epsilon} V_{\tau}$.} $\{ V_{\epsilon^{(i)}_{-}} \}_{i=1}^{k}$ near those points a sample can be \textit{exactly} reconstructed. The number $k$ of jumps over a bounded interval $[\epsilon_{1}, \epsilon_{2}]$ is itself a random variable and captures the memory needs of our approach.

Furthermore, Proposition \ref{thm:privateProcessJumpStatistics} suggests an efficient algorithm for directly sampling from the process $ \{ V_{\epsilon} \}_{\epsilon\in[\epsilon_{1},\epsilon_{2}]} $, which we present in Algorithm~\ref{alg:privateProcessSamplingL2}. Algorithm \ref{alg:privateProcessSamplingL2} draws a sample $\{ v_{\epsilon} \}_{\epsilon\in [\epsilon_{1},\epsilon_{2}]}$ from the stochastic process $V_{\epsilon}$ over a bounded interval $\epsilon\in[\epsilon_{1},\epsilon_{2}]$. This sample $\{v_{\epsilon}\}$ is the main object that performs diffusion of private data; whenever a user~$j$ requests user's $i$ private data $u_{i}$ residing $d_{ij}$ away, the estimator $y_{ij} = u_{i} + v_{\epsilon(d_{ij})}$.

The algorithm initializes a trace of the process by sampling from the Laplace distribution. Then, the algorithm extends this trace backwards in $\epsilon$ by sampling for the location of the next jump. The logarithm of the positions where jumps occur define a Poisson process with rate $\lambda = n+1$ and, thus, the length $\delta\epsilon = \ln\epsilon^{(i)} - \ln\epsilon^{(i+1)}$ of the interval until the next jump is exponentially distributed with density $\delta\epsilon \sim \lambda e^{-\lambda \, \delta\epsilon} $. Finally, conditioned on the event of a jump at $\epsilon^{(i)}$, the size $ \delta v = V_{\epsilon^{(i)}_{-}} - V_{\epsilon^{(i)}_{+}}$ of the jump is ``Bessel''- distributed with parameter $\frac{1}{\epsilon^{(i)}}$. The algorithm recycles until the level $\epsilon_{1}$ is reached. Additionally, responses $y_{ij}$ are generated upon request, and, thus, there is no excessive memory requirement $O(N^{2})$ for storing all the responses $\{ y_{ij} \}_{i,j \in \mathcal{V}}$. The number of iterations that Algorithm \ref{alg:privateProcessSamplingL2} performs is a random variable and is characterized by Proposition \ref{thm:privateProcessJumpStatistics}.

Typical single-dimensional ($n=1$) private data are date of birth, salary, and health status. For $n=2$, our results are applicable to \textit{geo-indistinguishability} \cite{andres2013geo} which is differential privacy for GPS locations and is experimentally illustrated in Subsection~\ref{sec:synthetic}. Finally, the case $n\to\infty$ appeals to private signals that appear in filtering problems and smart grid applications.

For completeness, Table~\ref{table:distributions} presents the parameterization of the elementary distributions used by the proposed algorithms. We note that the Bessel distribution decays exponentially and has closed-form expressions for odd $n$. Nonetheless, it is a single-dimensional distribution and, thus, discritization and sampling through the inverse cumulative function is possible.

\begin{algorithm}
	\begin{algorithmic}
		\Require{Dimension $n$; Privacy levels $\epsilon_{1}$ and $\epsilon_{2}$, such that $\epsilon_{2}>\epsilon_{1}>0$.}
		\Function{SamplePrivateProcessL2}{$n,\epsilon_{1},\epsilon_{2}$}
			\State $k\gets1$
			\State $\epsilon^{(1)} \gets \epsilon_{2}$
			\State $r \sim \text{Gamma}\left( T, \frac{1}{\epsilon_{2}} \right)$ \vspace{4pt}
			\State $v^{(1)}_{1},\ldots,v^{(1)}_{n} \overset{\text{i.i.d.}}{\sim} \text{Gaussian}(0,1)$ \vspace{4pt}
			\State $v^{(1)} \gets  \frac{ r }{ \| v^{(1)} \|_{2} } v^{(1)} $ \vspace{4pt}
			\While{$\epsilon(k) > \epsilon_{1}$}
				\State $\delta\epsilon \sim \text{Exponential}(n+1)$ \vspace{3pt}
				\State $\epsilon^{(k+1)} \gets e^{-\delta\epsilon} \, \epsilon^{(k)}$ \vspace{3pt}
				\State $r \sim \text{Bessel}\left(\frac{n}{2}-1, \frac{1}{\epsilon^{(k+1)}} \right)$ \vspace{4pt}
				\State $\delta v_{1}, \ldots, \delta v_{n} \overset{\text{i.i.d.}}{\sim} \text{Gaussian}(0,1)$ \vspace{3pt}
				\State $\delta v \gets \frac{r}{ \| \delta v \|_{2} } \delta v$ \vspace{4pt}
				\State $v^{(k+1)} \gets v^{(k)} + \delta v$
				\State $k \gets k+1$
			\EndWhile \vspace{2pt}
			\State Return $\{ (\epsilon^{(i)}, v^{(i)} \}_{i=1}^{k}$
		\EndFunction
	\end{algorithmic}
	\caption{Sampling from the stochastic process $V_{\epsilon}$ over a bounded interval $\epsilon\in[\epsilon_{1},\epsilon_{2}]$ can be performed both efficiently (with complexity $O\left( \ln \left( \frac{\epsilon_{2}}{\epsilon_{1}} \right) \right)$) and exactly (in the sense that we are not discretizing the interval or approximating the procssess).} \label{alg:privateProcessSamplingL2}
\end{algorithm}

\begin{table} \begin{center} \begin{tabular}{r||c|l|l}
\textbf{Distribution}	& \textbf{Param.}				& \textbf{Supp.}	& \textbf{Density}								\\ \hline
Laplace			& $\beta > 0$					& $x\in\reals$		& $ \frac{1}{2\beta} e^{-\frac{|x|}{\beta}} $ 	\\ \hline
Exponential		& $\lambda > 0$	 				& $x\in\reals_{+}$	& $ \lambda e^{-\lambda x} $					\\ \hline
Gamma			& \begin{tabular}{c} $n\in\mathbb{N}$, \\ $\beta > 0$ \end{tabular}	& $x\in\reals_{+}$	& $ \frac{1}{ \Gamma(n) \beta^{n} } x^{n-1} e^{-\frac{x}{\beta}} $	\\ \hline
Bessel			& \begin{tabular}{c} $n\in\mathbb{N}$, \\ $\beta > 0$ \end{tabular}	& $x\in\reals_{+}$	& \begin{tabular}{c} $ \frac{4}{\Gamma\left(\frac{n}{2}\right) (2\beta)^{\frac{n}{2}+1} } $ \\ $ x^{\frac{n}{2}} K_{\frac{n}{2}-1}\left( \frac{x}{\beta} \right) $ \end{tabular}
\end{tabular} \end{center} \caption{The distributions that are used by Algorithm~\ref{alg:privateProcessSamplingL2}. Sampling from these distributions can be performed using a uniform random variable and the quantile function.} \label{table:distributions} \end{table}

\section{Illustrative Examples} \label{sec:examples}
We present two application that depict diffusion of private data over a network. These example shows that bits of private information can be spread over the whole network, which allows users to estimate global quantities, such as epidemic spreading, while providing strong privacy guarantees.

\subsection{Synthetic Data} \label{sec:synthetic}
\begin{figure} \begin{center}
\includegraphics[width=\linewidth]{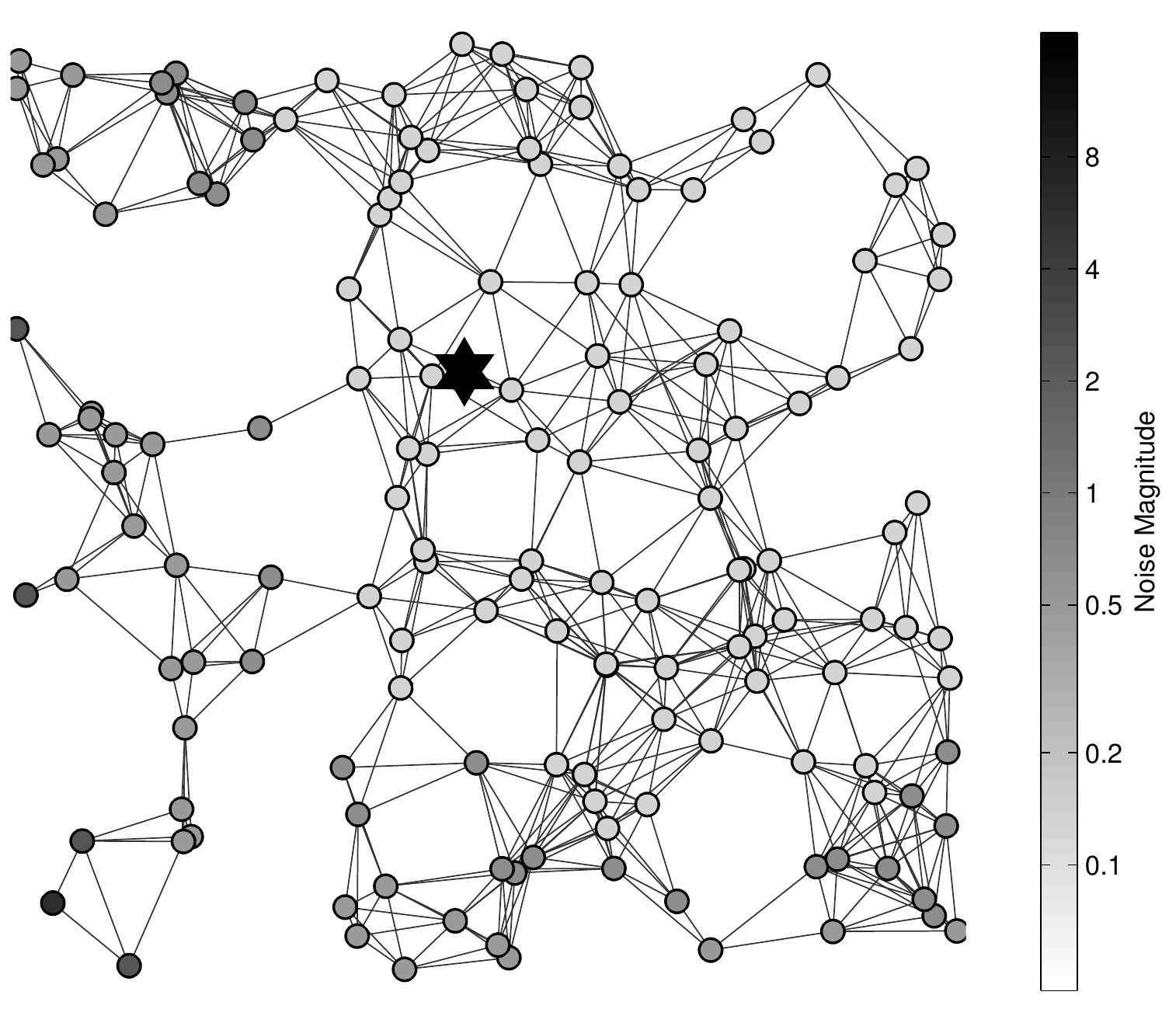}
\caption{Each individual $j$ gets the value $u_{i}+V_{\epsilon(d_{ij})}$, where $u_{i}$ is the true sensitive data, $d_{ij}$ is the number of hops between users $i$ and $j$, and $V_{\epsilon}$ is the result of Algorithm~\ref{alg:privateProcessSamplingL2}.} \label{fig:2DNetwork}
\end{center} \end{figure}

We consider the synthetic network in Figure \ref{fig:2DNetwork} with $N=|\mathcal{V}|=150$ nodes and $|\mathcal{E}|=1256$ edges, where edges are formed based on proximity. Each user $i\in\{1,\ldots,N\}$ wishes to publish her vector-valued private data $u_{i}\in\reals^{2}$, such as her GPS coordinates. For simplicity, we focus on a single user; our technique can be applied independently for each user. The distance $d_{ij}$ between users $i$ and $j$ is captured by the shortest path length. We choose an exponential function $\epsilon(\cdot)$ that converts distances \mbox{$d_{ij}\in\{1,\ldots,9\}$} to privacy levels $\epsilon(d_{ij}) \in [.5, 15]$.

\begin{figure} \begin{center}
\includegraphics[width=\linewidth]{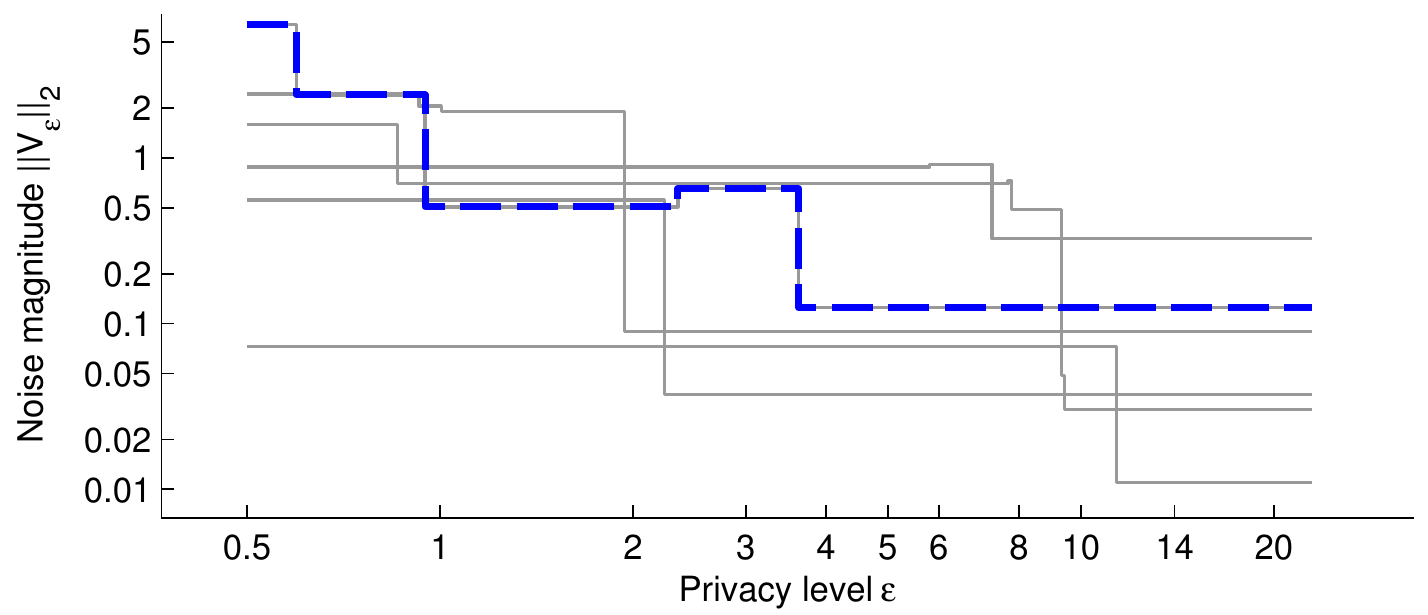}
\caption{Agent $i$ uses Algorithm \ref{alg:privateProcessSamplingL2} with $n=2$ and generates a single sample of the stochastic process. For small values of privacy level, high noise values are more likely, whereas, for loose privacy levels ($\epsilon\to\infty$), the noise values decrease in magnitude. Despite the continuity of the domain $\epsilon\in[0,\infty)$, the process performs only a few jumps.} \label{fig:2DProcess}
\end{center} \end{figure}

Algorithm \ref{alg:privateProcessSamplingL2} is executed by user $i$ for $n=2$ and the norms of several traces are shown in Figure \ref{fig:2DProcess}. For tight values of privacy level ($\epsilon\to0$), large amounts of noises are added. In Figure \ref{fig:2DNetwork}, nodes are colored based on the accuracy $\| y_{ij} - u_{i} \|_{2}$ of the response $y_{ij}$ they receive.

Although we have assumed the existence of a secure communication channel between any two users of the network and the existence of a central authority which computes the distances $d_{ij}$, an implementation that assumes only local communication between neighboring users and an \textit{honest-but-curious} model is possible. In such an implementation, user $i$ sends to all her neighbors the signal \mbox{$\{ u_{i} + V_{\epsilon} \}_{\epsilon\in(0,\epsilon(1))}$}. Then, each user $j$ receives the signal \mbox{$\{ u+V_{\epsilon} \}_{\epsilon\in[0, \epsilon(d_{ij}))}$}, trims it to \mbox{$\{ u+V_{\epsilon} \}_{\epsilon\in[0, \epsilon(d_{ij}+1))}$}, and broadcasts it to her friends.

\subsection{Real Dataset: Facebook}
In this section, we present an application of diffusing sensitive data on a real network. Specifically, an ``ego-network'' \cite{leskovec2012learning} is a the sub-graph $G=(\mathcal{V}\cup\{\text{Alice}\},\mathcal{E})$ of Facebook induced by a single user, Alice, and her friends~$\mathcal{V}$. Figure~\ref{fig:egoNetwork} plots such an ego-network, where the bottom-left node is the user whose neighborhood is captured. The rest of the nodes represent Alice's friends, edges represent friendships between her friends, whereas, the edges between Alice and her friends are omitted for clarity. We assume that Alice's infection status is captured by a single bit $u\in\{0,1\}$. Then, Alice wishes to share this information with her friends in a privacy-preserving way.

For each friend $i\in \mathcal{V}$, the distance $d_{i}$ is calculated by a central authority. Values $\{ d_{i} \}_{i\in \mathcal{V}}$ are independent of the private data $u$, and can be computed without any privacy requirements. In particular, values $d_{i}$ quantify the strength of the friendship between Alice and friend $i$ and are evaluated according to Equation \eqref{eqn:resistanceDistance}.
\begin{align} \label{eqn:resistanceDistance}
	d_{ij} = \Gamma_{ii} + \Gamma_{jj} - 2 \Gamma_{ij},
\end{align}
where $\Gamma\in\reals^{n\times n}$ is the pseudo-inverse of the Laplace matrix $L$ of the network. Due to space limitations, we use the fact that our technique allows post-processing of the responses $y_{ij}$ and, thus, is applicable for private bits.

Initially, Alice executes Algorithm \ref{alg:privateProcessSamplingL2} in order to generate a single sample $\{ w_{\epsilon} :\: \epsilon\in[ \underaccent{\bar}{\epsilon} ,\bar{\epsilon} ] \}$ of the stochastic process $\{ V_{\epsilon} : \: \epsilon>0 \}$, where $\underaccent{\bar}{\epsilon}$ (resp. $\bar{\epsilon}$) is a lower (resp. upper) bound of the quantity $\underset{i\in V}{\min} \: \epsilon(d_{i})$ (resp. $\underset{i\in V}{\max} \: \epsilon(d_{i})$). Function \mbox{$\epsilon(\cdot):\reals_{+}\to\reals_{+}$} is a decreasing function which converts distances $d_{i}$ to privacy levels $\epsilon_{i} = \epsilon(d_{i})$. In this example, we chose $\epsilon(d) = exp(-3.3 d + 4)$ which leads to privacy levels within $[.5,15]$. Next, individual responses are generated during run-time. Whenever user $i$ requests access to the sensitive data $u$, the response $y_{i}$ is securely communicated to user $i$:
\begin{align}
y_{i} = \Pi_{\{0,1\}} ( u + w_{\epsilon(d_{i})} ),
\end{align}
where $\Pi_{S}$ is the projection operator on the set $S$.

\begin{figure} \begin{center}
\includegraphics[width=\linewidth]{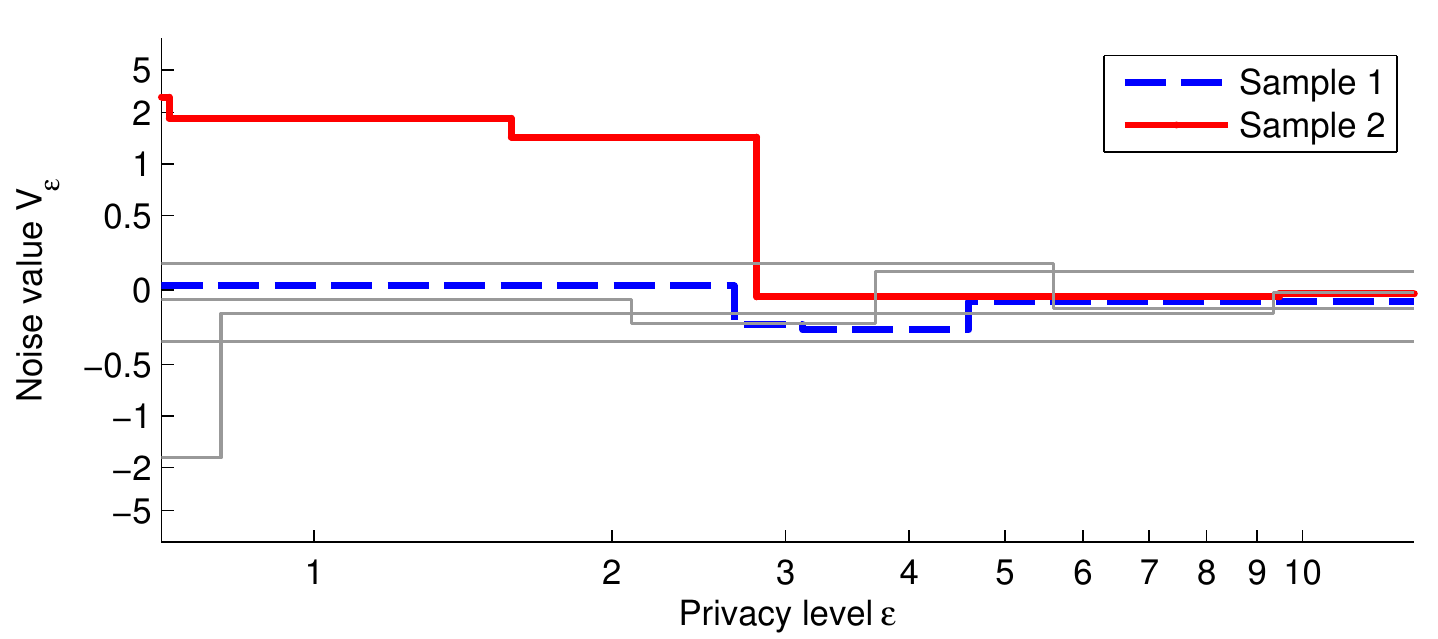}
\caption{Two samples of the stochastic process generated by Algorithm \ref{alg:privateProcessSamplingL2}. The samples are private information; a malicious user $i$ can subtract the noise $w_{\epsilon(d_{i})}$ from the received response $y_{i}$ and exactly infer the private data $u$.} \label{fig:egoProcess}
\end{center} \end{figure}

\begin{figure} \begin{center}
\includegraphics[width=\linewidth]{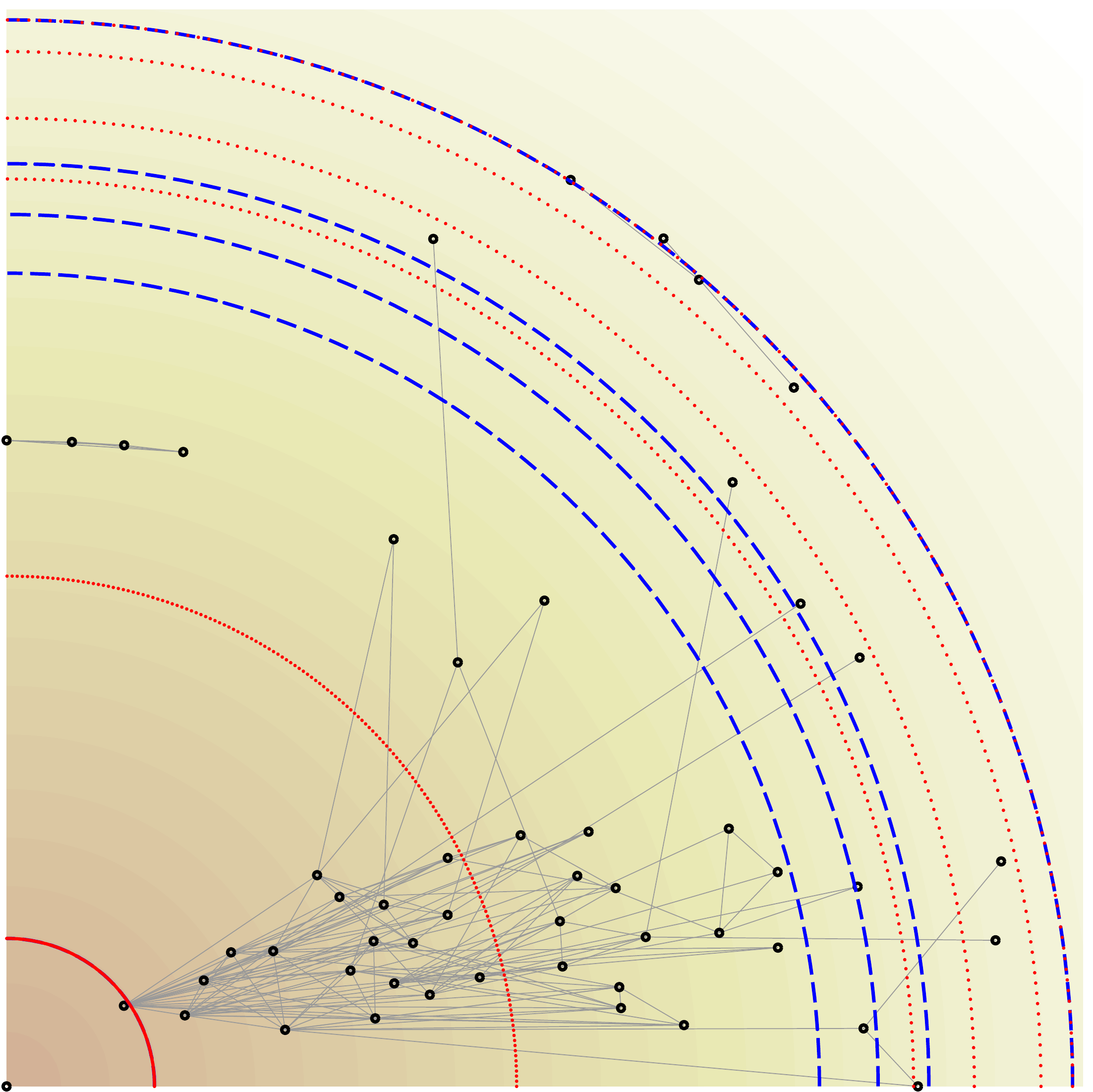}
\caption{An ego-network is the part of the Facebook network that is visible from a fixed user $A$ (ego), shown in the bottom-left corner of the plot. Each friend $i$ is plotted at distance $d_{i}$. The locations of the jumps of the two samples shown in Figure \ref{fig:egoProcess} are depicted by the blue and red circles. Although users residing within consecutive circles receive identical responses $y_{i}$, they are assigned different privacy levels $\epsilon(d_{i})$ and, thus, have different confidence levels.} \label{fig:egoNetwork}
\end{center} \end{figure}

Figure \ref{fig:egoProcess} depicts two executions of Algorithm \ref{alg:privateProcessSamplingL2} with \mbox{$n=1$}, whereas, Figure \ref{fig:egoNetwork} plots the ego-network centered around Alice. In particular, Alice is shown on the bottom-left corner and each friend $i$ is plotted at distance $d_{i}$ from her. The blue and red circles mark the jumps of the stochastic process for the two samples $w_{\epsilon}^{\text{blue}}$ and $w_{\epsilon}^{\text{red}}$. Counter-intuitively, friends $i$ lying within two consecutive blue circles receive \textit{exactly} the same response $y_{i}$ although they are assigned different privacy levels $\epsilon(d_{i})$. The paradox is settled by noticing that the boundary circles are random variables themselves. Therefore, users receiving identical responses have different confidence levels.

\section{Conclusions}
In this work, we considered the case of a network where each user owns a private data $u\in\reals^{n}$ such as her salary or her infection status and wishes to share approximations of this private data with the rest of the network under differential privacy guarantees. Specifically, we assumed that user $i$ requires $\epsilon(d_{ij})$-differential privacy against against user $j$, where $\epsilon(\cdot)$ is a decreasing function and $d_{ij}$ is the distance induced by the underlying network between users $i$ and $j$. In this context, we derived a composite mechanism that generates the response $y_{ij}$ as user's $j$ approximation of user's $i$ private data. The accuracy of the response $y_{ij}$ depends only on the allocated privacy level $\epsilon(d_{ij})$ and not on the size or other parameters of the network. An important property of our proposed mechanism is the resilience to coalitions where we considered a group of users combining their received responses for more accurate approximations. Practically, this means that scenarios where an adversarial user creates multiple fake accounts cannot weaken the privacy guarantees. Algorithms for sampling from this composite mechanism were also provided. In particular, the complexity of these algorithms is independent of the size of the network, which renders them scalable, and is dictated only by the extreme privacy levels $\min_{i\in\mathcal{V}}\epsilon(d_{ij})$ and $\max_{i\in\mathcal{V}}\epsilon(d_{ij})$. Finally, we provided two illustrative examples: one on a synthetic network where users communicate their private GPS locations, and one where a user shares her infection status with her Facebook ego-network. This work focused on the privacy aspect of the problem of diffusing private data over networks. Future work includes the joint problem of accurately estimating formally-defined global quantities while preserving privacy of users' data.

\appendix
\subsection{Proof of Theorem~\ref{thm:socialNetworkPrivacy}} \label{app:socialNetworkPrivacy}
Theorem~\ref{thm:socialNetworkPrivacy} is established in multiple steps. First, we focus on the discrete-domain process $\{ V_{\epsilon_{i}} \}_{i=1}^{m}$, where \mbox{$\epsilon_{1}\leq\dots\leq\epsilon_{m}$} and, in particular, on the case of $m=2$, with $\epsilon_{1} \leq \epsilon_{2} < \sqrt{2} \epsilon_{1}$, where the second inequality is due to technical reasons. Next, we prove the Markov property which allows $m$ discrete privacy levels. Finally, we pass to the limit and derive the continuous-domain process $\{ V_{\epsilon} \}_{\epsilon>0}$ as stated in Theorem~\ref{thm:socialNetworkPrivacy}.

\begin{proof}[Proof for two privacy levels]
	We consider the stochastic process $V_{\epsilon}$ supported on two privacy levels $\{ \epsilon_{1}, \epsilon_{2} \}$, where $\epsilon_{1} \leq \epsilon_{2} < \sqrt{2} \epsilon_{1}$. Allowing generalized functions, we assume that  the joint distribution of $V_{\epsilon_{1}}$ and $V_{\epsilon_{2}}$ has density:
	\begin{align} \label{eqn:L2GraduaPrivacy:1}
		\prob( V_{\epsilon_{1}} = x, V_{\epsilon_{2}} = y ) = l_{\epsilon_{1},\epsilon_{2}}(x,y) = g(x,y), \quad x,y\in\reals^{n}
	\end{align}
	Density \eqref{eqn:L2GraduaPrivacy:1} should satisfy the following marginal distributions and privacy constraints:
	\begin{align}
		\int_{\reals^{n}} g(x,y) d^{n}y & = \epsilon_{1}^{n} C_{1} e^{-\epsilon_{1} \|x\|_{2}}, \\
		\int_{\reals^{n}} g(x,y) d^{n}x & = \epsilon_{2}^{n} C_{1} e^{-\epsilon_{2} \|y\|_{2}}, \\
		\left\| \nabla_{x} g(x,y) + \nabla_{y} g(x,y) \right\|_{2} & \leq \epsilon_{2} g(x,y),
	\end{align}
	where $C_{1} = \frac{\Gamma(\frac{n}{2}+1) }{ \pi^{\frac{n}{2}} \Gamma(n+1) }$. The first two constraints express that $V_{\epsilon_{1}}$ and $V_{\epsilon_{2}}$ should be Laplace-distributed with parameters $\frac{1}{\epsilon_{1}}$ and $\frac{1}{\epsilon_{2}}$, respectively. The last constraint enforces that the mechanism that releases \mbox{$(u+V_{\epsilon_{1}}, u+V_{\epsilon_{2}})$} must be $\epsilon_{2}$-private. We solve for densities $g$ of the form
	\begin{align} \label{eqn:L2GradualPrivacy:4}
		g(x,y) = \epsilon_{2}^{n} C_{1} \phi(x-y) e^{-\epsilon_{2} \|y\|_{2}},
	\end{align}
	where $\phi:\reals^{n}\rightarrow\reals$ is a (possibly generalized) function satisfying
	\begin{align} \begin{split} \label{eqn:L2GradualPrivacy:2}
		\int_{\reals^{n}} \phi(x-u) \epsilon_{2}^{n} e^{-\epsilon_{2} \|u\|_{2}} d^{n}u & = \epsilon_{1}^{n} e^{-\epsilon_{1} \|x\|_{2}}, \\ 
		\int_{\reals^{n}} \phi(u) d^{n}u & = 1.
	\end{split} \end{align}
	The first equation in (\ref{eqn:L2GradualPrivacy:2}) is a $n$-dimensional convolution with solution
	\begin{align} \label{eqn:L2GradualPrivacy:3}
		\mathcal{F}\phi(s) = \frac{ \mathcal{M}(s;\epsilon_{1}) }{ \mathcal{M}(s;\epsilon_{2}) },
	\end{align}
	where $\mathcal{M}(s;\epsilon) = \mathcal{F}\left\{ \epsilon^{n} e^{-\epsilon \|x\|_{2}} \right\}(s)$, and $s\in\reals^{n}$ is the frequency. Solution \eqref{eqn:L2GradualPrivacy:3} satisfies the second equation in \eqref{eqn:L2GradualPrivacy:2} since
	\begin{align}
		\int_{\reals^{n}} \phi(u) d^{n}u = \left. \mathcal{F}\phi(s) \right|_{s=0} =  \frac{ \mathcal{M}(0;\epsilon_{1}) }{ \mathcal{M}(0;\epsilon_{2}) } = 1.
	\end{align}
	Finally, we need to prove that, for $\phi$ given in \eqref{eqn:L2GradualPrivacy:3}, density $g$ is well-defined, specifically:
	\begin{align}
		\phi(z) \geq 0, \quad \forall z\in\reals^{n}.
	\end{align}
	This is proven under the assumption that $\epsilon_{2} < \sqrt{2} \epsilon_{1}$; this assumption will eventually be removed. According to Lemma~\ref{lem:fourierPair:1}, we get:
	\begin{align} \begin{split} \label{eqn:L2GradualPrivacy:7}
		\mathcal{F}\phi(s)	& = \frac{\mathcal{M}(s;\epsilon_{1})}{\mathcal{M}(s;\epsilon_{2})} = \left( \frac{\epsilon_{1}}{\epsilon_{2}} \right)^{n+1} \left( 1 + \frac{\epsilon_{2}^{2}-\epsilon_{1}^{2}}{\epsilon_{1}^{2} + \rho^{2}} \right)^{\frac{n+1}{2}} \\
						& =\left( \frac{\epsilon_{1}}{\epsilon_{2}} \right)^{n+1} \sum_{k=0}^{\infty} \binom{\frac{n+1}{2}}{k} \left( \frac{ \frac{\epsilon_{2}^{2} }{ \epsilon_{1}^{2} } - 1}{1 + \frac{ \rho^{2} }{ \epsilon_{1}^{2} } } \right)^{k},
	\end{split} \end{align}
	where $\rho = \|s\|_{2}$. The sum in the right-hand side is an infinite series only when $n$ is even, and, for $\epsilon_{2}< \sqrt{2} \epsilon_{1}$, it converges uniformly in $s$ to the left-hand side. Lemma \ref{lem:fourierPair:2} can be used to invert the series:
	\begin{align} \begin{split} \label{eqn:L2GradualPrivacy:9}
		\phi(x) & = \left( \frac{\epsilon_{1}}{\epsilon_{2}} \right)^{n+1} \sum_{k=0}^{\infty} \binom{\frac{n+1}{2}}{k} *^{k} \Bigg\{ \left( \frac{\epsilon_{2}^{2} }{ \epsilon_{1}^{2} } - 1 \right) \epsilon_{1}^{n} (2\pi)^{-\frac{n}{2}} \\
		 & \qquad (\epsilon_{1} r)^{1-\frac{n}{2}} K_{\frac{n}{2}-1}(\epsilon_{1} r) \Bigg\},
	\end{split} \end{align}
	where $r = \|x\|_{2}$, $K_{k}(x)$ is the modified Bessel function of the second kind, and $*$ is the $n$-dimensional convolution. Since $\frac{\epsilon_{2}^{2}}{\epsilon_{1}^{2}} - 1\geq 0$ and $K_{\frac{n}{2}-1}(r)\geq 0$, density $g$ is well-defined.
\end{proof}

Next, we prove that the discrete-domain stochastic process $\{ V_{\epsilon_{i}} \}_{i\in\{1,\ldots,m\}}$ is Markov.
\begin{proof}[Proof of the Markov property]
	Consider the discrete-domain process $\{ V_{\epsilon_{i}} \}_{i\in\{1,\ldots,m\}}$ supported on $m$ non-decreasing privacy levels $\{\epsilon_{1},\ldots,\epsilon_{m}\}$, and the joint distribution that satisfies the Markov property:
	\begin{align} \begin{split} \label{eqn:L2GradualPrivacy:11}
		\prob( V_{\epsilon_{i}}=v_{i}, \: \forall i ) & = l_{\epsilon_{1:m}}(v_{1}, \ldots, v_{m}) \\
			& \hspace{-5pt} = \prob( V_{\epsilon_{1}}=v_{1} ) \prod_{i=2}^{m} \prob( V_{\epsilon_{i}}=v_{i} | V_{\epsilon_{i-1}}=v_{i-1} )  \\
			& =  l_{\epsilon_{1}}(v_{1}) \prod_{i=2}^{m} \frac{ l_{\epsilon_{i-1:i}}(v_{i-1},v_{i}) }{ l_{\epsilon_{i}}(v_{i}) },
	\end{split} \end{align}
	where $l_{\epsilon}(v) \propto e^{-\epsilon \|v\|_{2}}$ is the $n$-dimensional Laplace distribution with parameter $\epsilon^{-1}$ and $l_{\epsilon_{1},\epsilon_{2}}(v_{1},v_{2})$ is the density $g$ from the previous proof. Then, the joint distribution $l_{\epsilon_{1:m}}$ satisfies the following properties:
	\begin{itemize}
		\item \textit{Accuracy:} Each coordinate $V_{\epsilon_{i}}$ is optimally-distributed, i.e. Laplace-distributed with parameter $\epsilon_{i}^{-1}$:
			\begin{align}
				\prob(V_{\epsilon_{i}}=v_{k}) 	& = \int_{\reals^{n(m-1)}} l_{\epsilon_{1:m}}(v_{1}, \ldots, v_{m}) dv_{-i} \\
									& = l_{\epsilon_{i}}( v_{i} ),
			\end{align}
			where $dv_{-i} = dv_{1} \cdots dv_{i-1} \, dv_{i+1} \cdots dv_{m}$.
		\item \textit{Privacy:} The mechanism that releases $\{y_{i}\}_{i=1}^{m}$, where \mbox{$y_{i} = u + V_{\epsilon_{i}}$} is $\epsilon_{m}$-private. Indeed, the mechanism can be expressed as:
			\begin{align} \begin{split} \label{eqn:L2GradualPrivacy:12}
				\begin{bmatrix} y_{1} \\ \vdots \\ y_{m-1} \\ y_{m} \end{bmatrix} 	& = \begin{bmatrix} u+V_{\epsilon_{1}} \\ \vdots \\ u+V_{\epsilon_{m-1}} \\ u+V_{\epsilon_{m}} \end{bmatrix} \\
				& = (u+V_{\epsilon_{m}}) + \begin{bmatrix} \sum_{i=2}^{m} V_{\epsilon_{i-1}} - V_{\epsilon_{i}} \\ \vdots \\ V_{\epsilon_{m-1}} - V_{\epsilon_{m}} \\ 0 \end{bmatrix}.
			\end{split} \end{align}
			Density $l_{\epsilon_{i-1},\epsilon_{i}}$ defined in \eqref{eqn:L2GradualPrivacy:9} shows that $V_{\epsilon_{i-1}} - V_{\epsilon_{i}}$ is distributed independently of any other random variable. Thus, the mechanism can be viewed as the composition of the $\epsilon_{m}$-private mechanism that releases $u+V_{\epsilon_{m}}$ post-processed by adding independent noise. Since differential privacy is resilient to post-processing \cite{dwork2013algorithmic}, the overall mechanism is $\epsilon_{m}$-private.
	\end{itemize}
\end{proof}

Finally, we derive the continuous domain process $\{ V_{\epsilon} \}_{\epsilon>0}$ by passing to the limit as the $m\to\infty$, $\epsilon_{1}=0$, and $\epsilon\to\infty$. Specifically, we derive closed-form expressions that lead to efficient algorithms for sampling of the continuous-domain stochastic process.

\begin{proof}[Proof of the continuous-domain process]
	In density~\eqref{eqn:L2GradualPrivacy:9}, let $\epsilon_{1} = \epsilon$ and $\epsilon_{2} = (1+\delta) \epsilon$, where $0 < \delta \ll 1$. Then, we prove that we can safely ignore high-order terms:
	\begin{align} \label{eqn:L2GradualPrivacy:10}
		\phi_{\epsilon}(x)		& \propto \delta(x) + \mathcal{F}^{-1} \left\{ \frac{(n+1)\delta}{1+\frac{\rho^{2}}{\epsilon^{2}}} \right\} + O\left(\delta^{2}\right) \\
					& = \delta(x) + \frac{\epsilon^{n} (n+1)}{(2\pi)^{\frac{n}{2}}} (\epsilon r)^{1-\frac{n}{2}} K_{\frac{n}{2}-1}(\epsilon r) \delta + O\left(\delta^{2}\right),
	\end{align}
	where $r=\|x\|_{2}$. We discretize a bounded interval $[\underline{\epsilon}, \overline{\epsilon}]$ by considering $K+1$ points $\epsilon^{(i)} = q^{i} \underline{\epsilon}$, where $q = \left( \frac{\overline{\epsilon}}{\underline{\epsilon}} \right)^{K^{-1}}$, and define the random variable $Z$ as follows:
	\begin{align}
		Z := V_{\underline{\epsilon}} - V_{\overline{\epsilon}} = \sum_{i=1}^{K} V_{\epsilon^{(i-1)}} - V_{\epsilon^{(i)}},
	\end{align}
	where the random variables $\{ V_{\epsilon^{(i)}} \}_{i=0}^{K}$ form a discrete-domain stochastic process introduced in \eqref{eqn:L2GradualPrivacy:11}. For large $K$, the step $\delta = q-1$ becomes arbitrarily small and, thus, we use the first-order approximation in \ref{eqn:L2GradualPrivacy:10} for each telescoping term $(V_{\epsilon^{(i-1)}} - V_{\epsilon^{(i)}}) \sim \phi_{\epsilon^{(i)}}$. Finally, the random variable $Z$ is distributed as:
	\begin{align}
		Z	& \sim *_{i=1}^{N} \phi_{\epsilon^{(i)}}(Z) \\
			& = *_{i=1}^{N} \Bigg\{ \delta(Z) + \frac{ \left( \epsilon^{(i)} \right)^{n} (n+1)}{(2\pi)^{\frac{n}{2}}} (\epsilon^{(i)} \|Z\|_{2})^{1-\frac{n}{2}} \\
			& \qquad K_{\frac{n}{2}-1}(\epsilon^{(i)} \|Z\|_{2}) \delta \Bigg\} + O\left(\delta\right),
	\end{align}
	where we let $\delta\to0$. This proves that we can approximate the continuous-domain stochastic process by a first-order approximation of the discrete-domain process.
\end{proof}

Equation~\eqref{eqn:L2GradualPrivacy:10} characterizes the stochastic process $\{ V_{\epsilon} \}_{\epsilon>0}$. The atom renders the stochastic process lazy; with high probability, the process is constant over sufficiently small intervals. The linear term governs the statistics of the the jump.

\subsection{Proof of Proposition~\ref{thm:privateProcessJumpStatistics}}
We now provide the proof of Proposition~\ref{thm:privateProcessJumpStatistics} that characterizes the jumps of the stochastic process $\{ V_{\epsilon} \}_{\epsilon>0}$ and, thus, captures the complexity of Algorithm~\ref{alg:privateProcessSamplingL2}.

\begin{proof}
Consider the first-order approximation of the backwards conditional distribution $\phi_{\epsilon}$ derived in \eqref{eqn:L2GradualPrivacy:10}, where $0 < \epsilon$ and $0<\delta\ll 1$:
	\begin{align} \label{eqn:privateProcessJumpStatistics:2}
		\prob( V_{\epsilon} = x | V_{(1+\delta)\epsilon} = y ) \approx \left( 1 + (n+1)\delta \right)^{-1}  \\
			\quad \Big( \delta(x) + \frac{\epsilon^{n} (n+1)}{(2\pi)^{\frac{n}{2}}} (\epsilon r)^{1-\frac{n}{2}} K_{\frac{n}{2}-1}(\epsilon r) \delta \Big)
	\end{align}
	Let $a_{n}(x)$ denote the probability that the process performs $n$ jumps in the interval $[\epsilon, e^{x}\epsilon]$. According to \eqref{eqn:privateProcessJumpStatistics:2} shows that, for sufficiently small intervals $[\epsilon, (1+\delta)\epsilon]$, the process remains constant with probability $\left( 1 + (n+1)\delta \right)^{-1}$, therefore, $a_{n}(x)$ is invariant of $\epsilon$. Under the discretization introduced earlier, where $\underline{\epsilon}\leftarrow\epsilon$ and $\overline{\epsilon}\leftarrow e^{x} \epsilon$:
	\begin{align}
		a_{0}(x) = \prob( 0 \text{ jumps in } [\epsilon, e^{x}\epsilon] ) = e^{-(n+1)x}.
	\end{align}
	A limiting argument is used to compute $a_{1}(x)$:
	\begin{align}
		a_{1}(x)	& = \lim_{K\to\infty} \sum_{k=1}^{K} \prob\left( 0 \text{ jumps in } [\epsilon,\epsilon^{(k-1)}] \right) \\
					& \prob\left( 1 \text{ jump in } [\epsilon^{(k-1)},\epsilon^{(k)}] \right)  \prob\left( 0 \text{ jumps in } [\epsilon^{(k)}, e^{x} \epsilon] \right) \\
					& = (n+1)x e^{-(n+1)x}.
	\end{align}
	A similar argument provides a recurrent equation and eventually:
	\begin{align} \label{eqn:privateProcessJumpStatistics:3}
		a_{k}(x) = \frac{ ((n+1)x)^{k} }{k!} e^{-(n+1)x}.
	\end{align}
	Therefore, for a bounded interval interval $[\underline{\epsilon}, \overline{\epsilon}]$, the number $n$ of jumps is characterized by distribution \eqref{eqn:privateProcessJumpStatistics:3}, which is the Poisson distribution with mean value $(n+1) \ln \left( \frac{ \overline{\epsilon} }{ \underline{\epsilon} } \right) $.
\end{proof}

\subsection{Fourier Transform Pairs}
In this section, we derive two Fourier pairs used in the proof of Theorem \ref{thm:socialNetworkPrivacy}. By convention, the following definition of Fourier transform $f \overset{ \mathcal{F} }{\leftrightarrow} F$ is used:
\begin{align} \label{eqn:fourierLemmas:1}
\mathcal{F}\left\{ f(x) \right\}(s) 		& = \int_{\reals^{n}} f(x) e^{-j x \cdot s } d^{n}x, 		 \\
\end{align}
where $f,F: \reals^{n}\rightarrow \reals$.

\begin{lemma} \label{lem:fourierPair:1}
The $n$-dimensional Fourier transform $\mathcal{F}$ of $f:\reals^{n}\rightarrow\reals$:
\begin{align}
	f(x)=e^{-\|x\|_{2}}
\end{align}
is:
\begin{align}
\mathcal{F}\left\{ f(x) \right\}(s) = \frac{ \pi^{\frac{n}{2}}  \Gamma(n+1) }{ \Gamma\left(\frac{n}{2}+1 \right) } \left( 1 + \|s\|_{2}^{2} \right)^{-\frac{n+1}{2}},
\end{align}
where $s\in\reals^{n}$.
\end{lemma}
\begin{lemma} \label{lem:fourierPair:2}
The $n$-dimensional Fourier transform $\mathcal{F}$ of $f:\reals^{n}\rightarrow\reals$, $f(x)=\|x\|^{1-\frac{n}{2}} K_{\frac{n}{2}-1}(\|x\|)$, is:
\begin{align}
\mathcal{F}\left\{ f(x) \right\}(s) =  \frac{ (2\pi)^{\frac{n}{2}} }{1+\rho^{2}}, \\
\end{align}
where $x\in\reals^{n}$, $\rho=\|s\|_{2}$, and $K_{k}(x)$ is the modified Bessel function of the second kind.
\end{lemma}
	The integrals are formulated using spherical coordinates and, then, symbolically evaluated with Mathematica 10.0. For an non-automated evaluation of the expressions, we refer the reader to MathWorld \cite{mathworld:hypergeometric} and references therein, and integral look-up tables \cite{abramowitz1964handbook}.


\bibliographystyle{IEEEtran}
\bibliography{privacy}

\end{document}